\documentclass{cseg2019}
\usepackage{xy}
\usepackage{amsmath}
\usepackage{graphicx}
\usepackage{natbib}
\usepackage{supertabular}
\usepackage{algpseudocode}
\usepackage{algorithm}
\usepackage{accents}
\usepackage{epstopdf}
\usepackage{rotating}
\usepackage{lscape}
\usepackage[caption=false]{subfig}
\usepackage [colorlinks=true, citecolor=black, linkcolor=black]{hyperref}

\begin{document}
 \vspace*{0cm}
\begin{center}

 \LARGE {\bf Least-squares RTM of a seismic-while-drilling dataset}\\
\vspace{10pt} \small \color{black}Nasser Kazemi, Daniel Trad, Kris Innanen, Roman Shor\\
\vspace{10pt}  University of Calgary, Calgary, Canada \\

\end{center}

\begin{abstract}
Least-squares migration can, in theory, reduce the acquisition footprint and improve the illumination of the subsurface structures. It can also recover the amplitudes of the events to some extent. However, the migration operator is not complete. In other words, the operator does not span the full range of the model and the portion of the model that is in the null space of the operator will not be recovered even by posing imaging as an inverse problem. In geophysical terminology, in complex subsurface structures, rays or the wave energy will penetrate poorly in some regions, e.g., subsalt region, and that region will be a shadow zone to our acquisition system. The shadow zone is in the null space of the migration operator and the subsurface information in that region will not be recovered. Accordingly, in this research, we aim at using another set of data whose ray paths are different from the surface seismic. Seismic-while-drilling (SWD) dataset are complementary to surface data, and it brings an opportunity to address seismic illumination issue by adding new measurements into the imaging problem. After considering the correlative and non-impulsive nature of the SWD source signature, the prestack least-squares depth migration of the SWD dataset can be achieved. We study the feasibility of the least-squares reverse time migration of the SWD dataset and its potential in imaging the parts of the model which are in the shadow zone of the surface seismic acquisition.
\end{abstract}
\section{Introduction}
\vspace{-.3cm}
In conventional surface seismic acquisition, back propagating the recorded wave fields at the surface through a background medium can result in imaging the subsurface structures. This process, called seismic imaging, suffers from non-uniform illuminations.  The imaging algorithms can usually address the kinematics of the subsurface structures adequately. However, they fail to properly provide the amplitude information of the events. To remedy this shortcoming, authors pose the seismic imaging as an inverse problem. Least-squares migration can be used for this purpose. Least-squares migration with Kirchhoff operator was one of the early attempts \cite[]{tarantola1984linearized,nemeth1999least,trad2015least}. The algorithm is also implemented with one-way wave equation operators \cite[]{Kuhl, clapp2005regularized,kazemi2015block}. To fully account for all kind of dips and complexities of the subsurface structure, authors introduce two-way propagators such as reverse time migration \cite[]{baysal1983reverse,loewenthal1983reversed,levin1984principle} into the least-squares imaging \cite[]{ji2009exact,wong2011least,dai2013plane,zhang2014stable, xue2015seismic,xu2017preconditioned,chen2017elastic}. Least-squares reverse time migration can, in theory, reduce the acquisition footprint and improve the illumination of the subsurface structures. It can also recover the amplitudes of the events to some extent. However, in the complex subsurface structure such as subsalt regions, waves penetrate weekly, and the reflected wave fields do not contain information from those regions. In other words, these regions will be in the shadow zone of the typical aperture-limited surface seismic acquisition. Accordingly, the amplitude information in the shadow zones will not be recovered even by posing imaging as an inverse problem. 

By contrast, in seismic-while-drilling (SWD) acquisition, drill bit-rock interaction can radiate significant elastic energy into the medium of interest which are strong enough to reach the surface and recorded by active receivers. Hence, the drill bit-rock interaction can be used as a seismic source. Moreover, having access to new source positions close to the target region will generate unique ray paths that are different from surface seismic. Provided that we understand the radiation patterns of the SWD sources, the SWD data contains information from within the medium of interest, at points normally not available to seismic sources. In this writeup, we explore the possibility of improving the subsurface illumination by implementing least-squares reverse time migration on a synthetic SWD dataset.        

\section{Least-squares reverse time migration}
\vspace{-.3cm}
Assuming a shot independent reflectivity model of the subsurface ${\bf m}$, the born forward modelling can be expressed as
\begin{equation}\label{T20}
{\bf d}={\bf L}{\bf m},\vspace{-.15cm}
\end{equation}
where ${\bf d}$ is the forward modelled data, and ${\bf L}$ is the forward modelling or de-migration operator. Application of the adjoint operator on the forward modelled data results in the conventional migration 
\begin{equation}\label{T21}
{\bf m}_{adj}={\bf L}^T{\bf d},\vspace{-.15cm}
\end{equation}
where ${\bf L}^T$ is the reverse time migration operator. The adjoint migrated image has poor illumination and suffers from amplitude biasing. To improve illumination and balance the amplitudes, one can pose imaging as an inverse problem. Posing imaging as an inverse problem not only improves the quality of the images but also provides an opportunity to include regularization and prior information about the subsurface. By doing so, one can emphasize good features in the final image. Here, we pose imaging as a least-squares minimization 
\begin{equation}\label{T22} 
 {{\bf m}_{LS}}=\underset{{\bf m}}{\operatorname{argmin}}\; \Arrowvert {\bf L}{\bf m}-{\bf d}\Arrowvert_2^2+\mu {\cal R}({\bf m}),\vspace{-.15cm}
\end{equation}
where ${\cal R}({\bf m})$ is a regularization term that enhances desired features in the model, and $\mu$ is a regularization parameter that balances the importance of data fidelity versus the regularization term.  We assume that the exact subsurface model has continuous and smooth features in spatial directions. One possible cost function that minimizes the data misfit and promotes smoothness in the model, is
 \begin{equation}\label{T23}
{{\bf m}_{LS}}=\underset{{\bf m}}{\operatorname{argmin}}\; \Arrowvert {\bf L}{\bf m}-{\bf d}\Arrowvert_2^2+\mu \Arrowvert {\bf D}{\bf m}\Arrowvert_2^2,\vspace{-.15cm}
\end{equation}
where ${\bf D}$ is a second order derivative operator. Equation \ref{T23} has a closed form solution
\begin{equation}\label{T24}
{{\bf m}_{LS}}=({\bf L}^T{\bf L}+\mu {\bf D}^T{\bf D})^{-1}{\bf L}^T{\bf d}.\vspace{-.15cm}
\end{equation}
Equation \ref{T23} can also be written as a typical least-squares minimization problem by concatenating the operators
\begin{equation}\label{T25}
{\bf y}_{LS}=\underset{{\bf y}}{\operatorname{argmin}}\; \Arrowvert \begin{bmatrix} {\bf{L}}{\bf D}^{-1} \\ \sqrt \mu\;{\bf I}\end{bmatrix} {\bf y}- \begin{bmatrix} {\bf d} \\ {\bf 0}\end{bmatrix}||_2^2=||\underaccent{\tilde}{{\bf A}}{\bf y}-{\bf b}||_2^2,\vspace{-.15cm}
\end{equation} 
where $\underaccent{\tilde}{{\bf A}}=\begin{bmatrix} {\bf{L}}{\bf D}^{-1} \\ \sqrt \mu\;{\bf I}\end{bmatrix}$ and ${\bf b}=\begin{bmatrix} {\bf d} \\ {\bf 0}\end{bmatrix}$. Note that ${\bf m}_{LS}={\bf D}^{-1}{\bf y}_{LS}$. In real-world applications, least-squares migration is considered as a medium to a large-scale problem and requires the application of iterative algorithms.  We use Conjugate Gradient algorithm to solve equation \ref{T25}. The pseudocode of the Conjugate Gradient algorithm for solving equation \ref{T25} is represented in Algorithm \ref{alg:1}. To build the forward and adjoint operators for prestack depth migration of SWD dataset, we need to estimate the source signature from the data. The wavelet estimation is done by using Sparse Multichannel Blind Deconvolution (SMBD) algorithm \cite[]{kazemi2014sparse}. 
\begin{algorithm}
 \caption{Conjugate Gradient algorithm with regularization}
 \label{alg:1}
  \begin{algorithmic}
 \State choose ${\bf y}_0$, 
 
 \State ${\bf s}_0={{\bf b}}-\underaccent{\tilde}{{\bf A}}{\bf y}_0$,
 
 \State ${\bf r}_0={\bf p}_0=\underaccent{\tilde}{{\bf A}}^{T}({\bf b}-\underaccent{\tilde}{{\bf A}}{\bf y}_0)$,
 
 \State ${\bf q}_0=\underaccent{\tilde}{{\bf A}}{\bf p}_0$,
 
 \State Initialize iteration $k=0$,
 \While {($||\underaccent{\tilde}{{\bf A}}{\bf y}_k-{{\bf b}}||_2^2>tol$)},
 
 \State $\alpha_{k+1}=<{\bf r}_k,{\bf r}_k>/<{\bf q}_{k},{\bf q}_{k}>$,
 
 \State ${\bf y}_{k+1}={\bf y}_k+\alpha_{k+1}\;{\bf p}_k$,
 
 \State ${\bf s}_{k+1}={\bf s}_k-\alpha_{k+1}\;{\bf q}_k$,
 
 \State ${\bf r}_{k+1}=\underaccent{\tilde}{{\bf A}}^{T}{\bf s}_{k+1}$,
 
 \State $\beta_{k+1}=<{\bf r}_{k+1},{\bf r}_{k+1}>/<{\bf r}_{k},{\bf r}_{k}>$,
 
 \State ${\bf p}_{k+1}={\bf r}_{k+1}+\beta_{k+1}{\bf p}_k$,
 
 \State ${\bf q}_{k+1}=\underaccent{\tilde}{{\bf A}}{\bf p}_{k+1}$,
 \State $k\leftarrow k+1$,
 \EndWhile
 \State ${\bf y}\leftarrow{\bf y}_k$, Calculate ${\bf m}={\bf D}^{-1}{\bf y}$.
  \end {algorithmic}
  
\end{algorithm}
\vspace{-0.2cm}
\section{Examples}
\begin{figure}[t]%
 \vspace{-.4cm}
\centering  
\parbox{3in}{ \centering 
   \includegraphics[width=.5\textwidth, height=7cm]{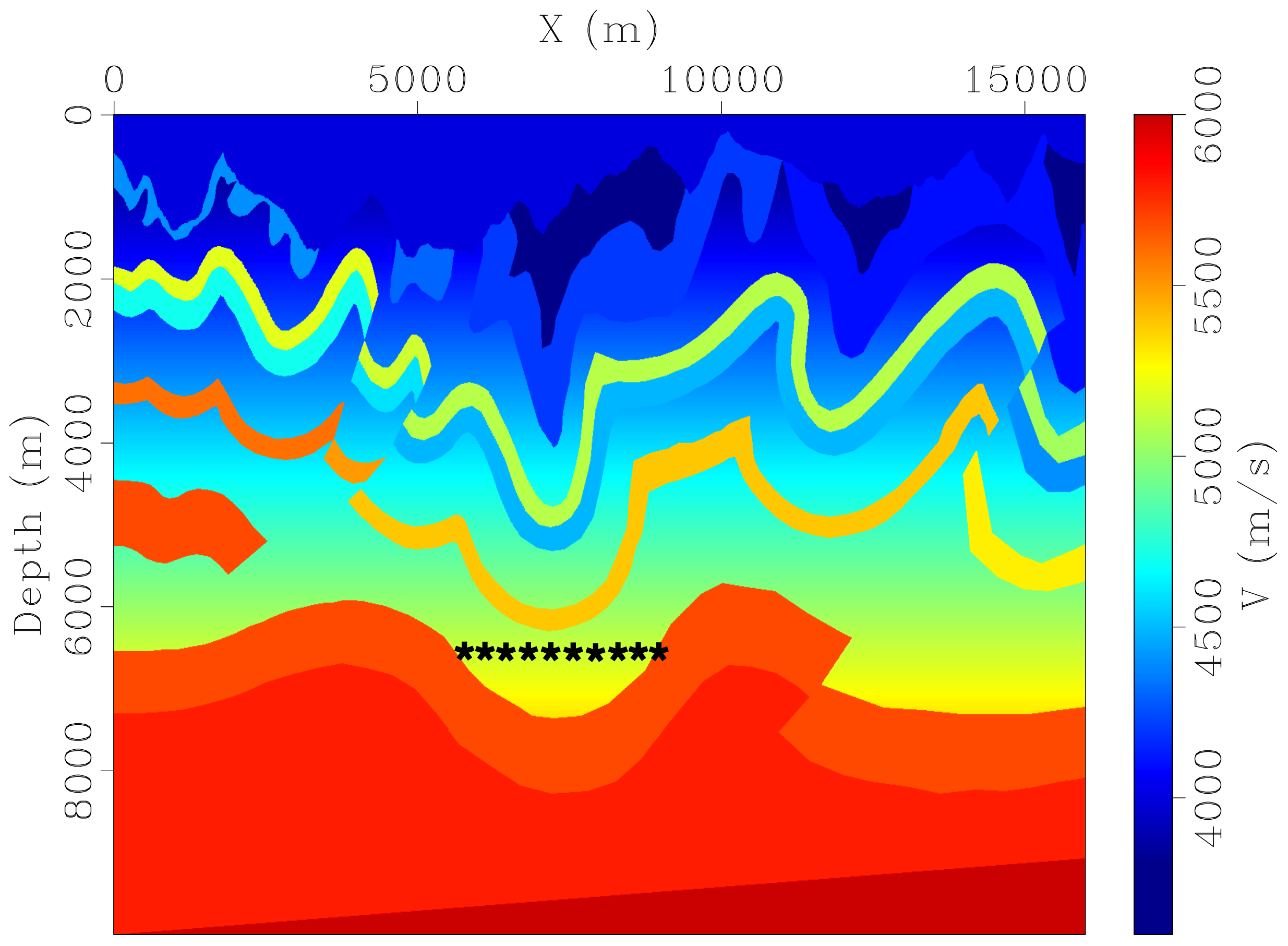}
   \vspace{-.9cm}  
   \caption{SWD acquisition geometry over BP velocity model. Black stars in the middle of the model around $6.4 km$ are SWD sources.}
\label{fig1}}%
\qquad
\parbox{3in}{ \centering
   \includegraphics[width=.43\textwidth, height=7cm]{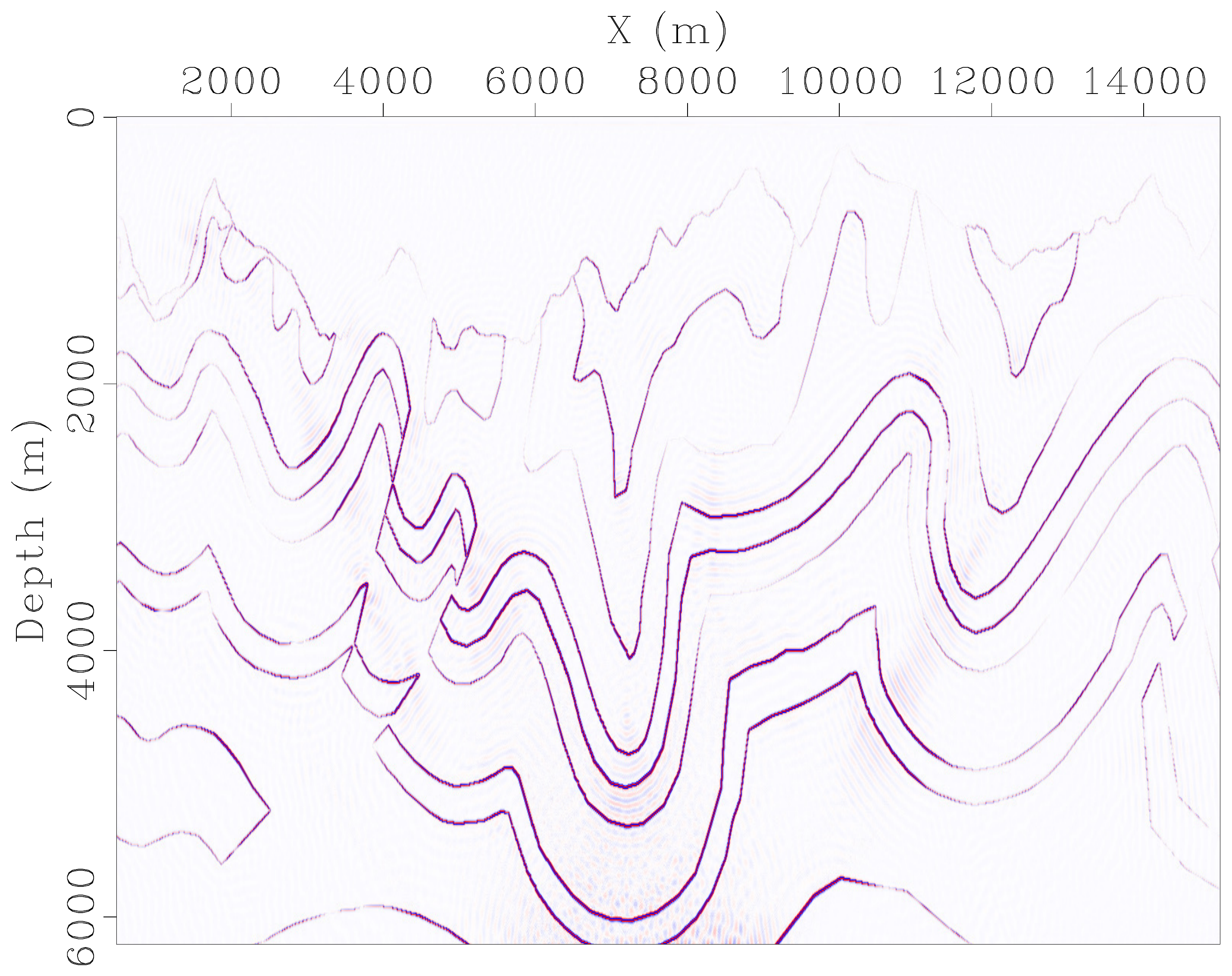}  
   \vspace{-0.1cm}
   \caption{Least-squares reverse time migration of the SWD dataset after $10$ iterations.}
  \label{fig2}}%
  \vspace{-.5cm}
\end{figure}

\begin{figure}[h]
 \vspace{-0.1cm}
\centering  
\parbox{3in}{ \centering 
   \includegraphics[width=.45\textwidth, height=6cm]{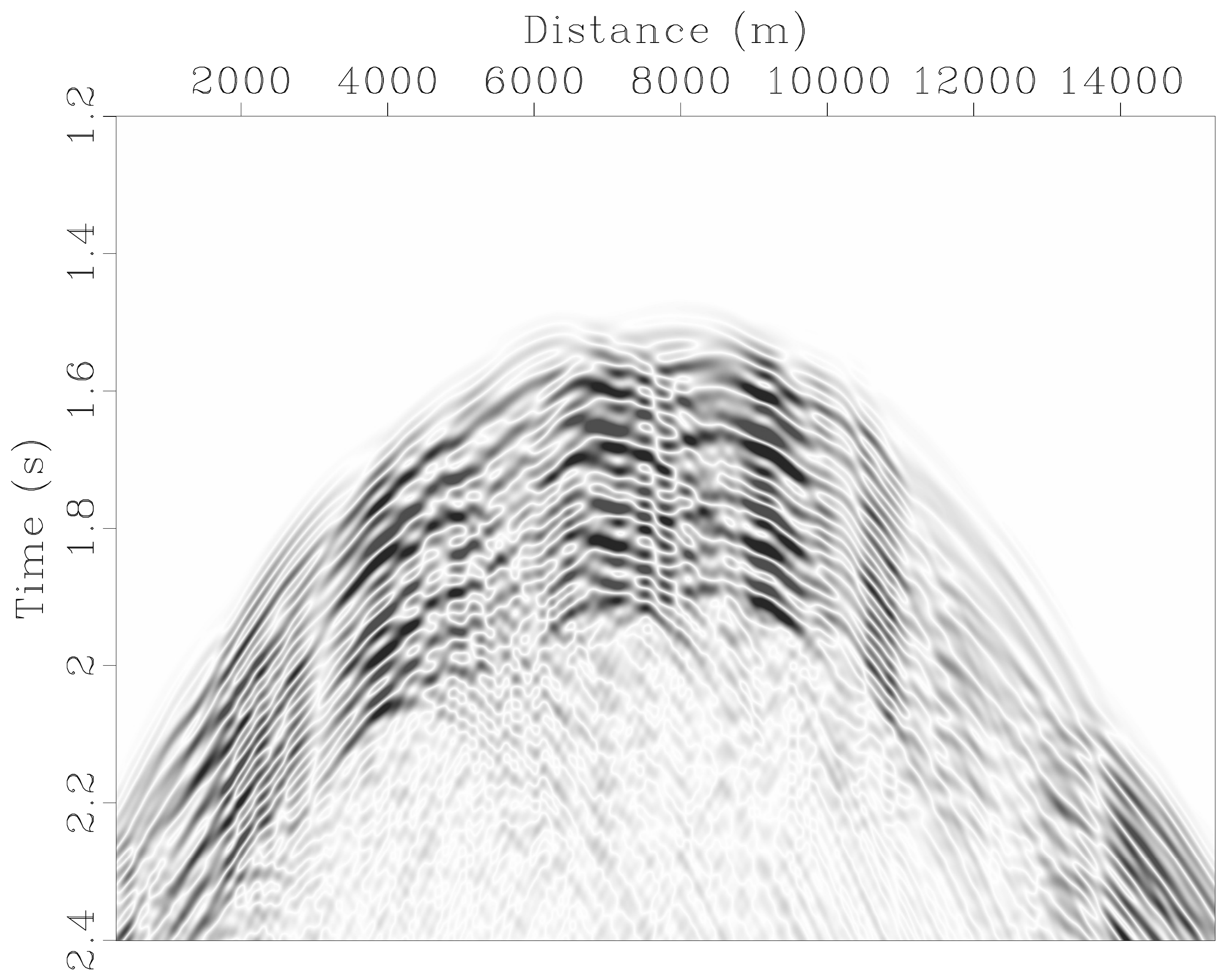}
  \vspace{-0.85cm}  
   \caption{True SWD shot gather corresponding to the fifth SWD source.}
\label{fig3}}%
\qquad
\parbox{3in}{ \centering
   \includegraphics[width=.45\textwidth, height=6cm]{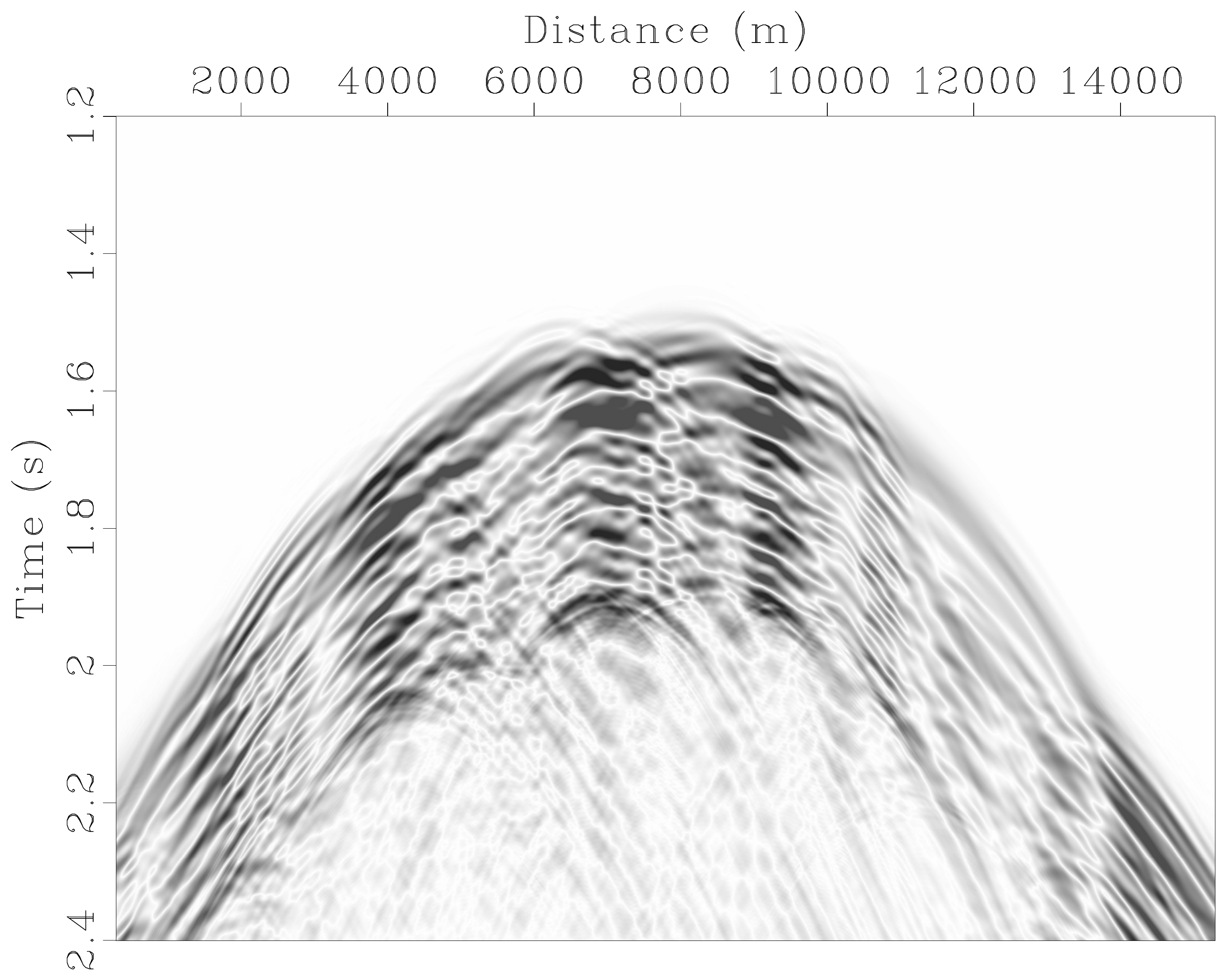}  
   \vspace{-0.85cm}
   \caption{Predicted fifth SWD shot gather using the inverted image shown in Figure \ref{fig2}.}
  \label{fig4}}%
\end{figure}

\vspace{-.3cm}
To evaluate the performance of the least-squares reverse time migration of a SWD dataset, we use a BP model represented in Figure \ref{fig1}. The drill bit- rock interaction is used as a seismic source in the deeper part of the model. The SWD acquisition geometry consists of $10$ SWD sources around $6.4km$, and the receivers are listening at the surface. There are $1500$ receivers at the surface with $10m$ intervals, and the source spacing is $300m$. To simulate the data, we use acoustic finite difference modelling and then convolve the data with a drill bit source signature. The drill bit-rock signature is similar to the work of \cite{kazemi2018illumination} and follows the assumptions of \cite{Poletto}. 
The SWD data is different from the surface seismic data. The SWD source signature is correlative and non-impulsive, and this makes it difficult to track the events. Hence, pre-processing the data and estimating the SWD source signature is an important step in our least-squares migration algorithm. After feeding the raw SWD shot gathers as receiver side wave field and the estimated SWD source signature for building the source side wave field, we achieve the least-squares migrated image (Figure \ref{fig2}). We use $10$ iterations of the Conjugate Gradient represented in Algorithm \ref{alg:1}. The result shows that the least-squares reverse time migration of the SWD dataset successfully imaged the subsurface structures. Moreover, finding a model that can predict the recorded data is one of the advantages of the least-squares algorithm over the conventional migration techniques. Figure \ref{fig3} shows the true shot gather of the SWD dataset and the predicted SWD data after applying the forward modelling engine ${\bf L}$ on the least-squares migrated image ${\bf m}$ is shown in Figure \ref{fig4}. The result shows that the least-squares method fits the recorded SWD data. 

\vspace{0.1cm}
\section{Conclusions}
\vspace{-.25cm}
We explored the possibility of using a drill bit-rock interaction as a seismic source. Moreover, we showed that SWD data contain valuable information from the subsurface, and, thanks to the unique ray paths of SWD acquisition, the SWD data bring an opportunity to mitigate non-uniform subsurface illumination. We formulated the least-squares reverse time migration of a SWD dataset and successfully imaged the subsurface structures. The crucial step in imaging the subsurface by using a SWD dataset is that we understand the radiation patterns of drill bit-rock interactions and be able to estimate the correlative and non-impulsive SWD source from the dataset. SWD source signature is one of the main inputs of our current workflow. To estimate the SWD source signature, we implemented a sparse multichannel blind deconvolution (SMBD) algorithm. Our preliminary results on the BP model showed that SWD data have the potential of imaging the subsurface structures. The next step is to combine surface seismic data with SWD data and apply a joint least-squares migration of combined dataset.
\vspace{-.3cm}
\section{Acknowledgment}
\vspace{-.3cm}
This research was undertaken thanks in part to funding from the Canada First Research Excellence Fund at the University of Calgary. The work was also funded by CREWES industrial sponsors.
\vspace{-.3cm}
\bibliography{CSEG-2019_final}

\end{document}